\documentclass[aps,pra,twocolumn,showpacs]{revtex4}
\usepackage{amsmath,amssymb,graphicx,bbold}
\usepackage{graphicx}  % needed for figures
\usepackage{dcolumn}   % needed for some tables
\usepackage{bm}        % for math
\usepackage{color}

\providecommand{\moy}[1]{\langle #1 \rangle}

\providecommand{\ket}[1]{\lvert #1 \rangle}

\begin{document}

\title{Bell inequality violation and operator ordering in quantum theory}
\author{H. M. Faria, K. Dechoum and A. Z. Khoury}

\affiliation{Instituto de F\'\i sica, Universidade Federal Fluminense,
24210-346 Niter\'oi, RJ, Brasil}

\date{\today}

\begin{abstract}
We investigate the role played by quantum operator ordering in the correlations that characterize two-photon polarization Bell measurements. 
The Clauser-Horne-Shimony-Holt (CHSH) criterion is investigated in the normal ordering imposed by the photodetection theory 
and in the symmetric ordering that constitutes the standard prescription for building Hermitian operators from products of 
non commuting observables. The two approaches are obtained in a single theoretical framework, where operator 
ordering is directly associated with the representation used for the density matrix. Moreover, this discussion can be recast 
in terms of the contribution given by the vacuum fluctuations to the detected signals. We also envisage possible detection schemes sensitive 
to these fluctuations with recent technological developments.
\end{abstract}
\pacs{03.65.Ud, 03.67.Mn, 42.50.Dv}
%\vskip2pc 

\maketitle

\section{introduction}
\label{intro}

Bell inequality violation has been used as a key criterion for identifying nonlocal or noncotextual correlations
in quantum mechanics 
\cite{Freedman1972,Bell-Aspect,bell_aspect_2004,Kwiat-Steinberg_1993,Kwiat-Steinberg_1994,Zeilinger1998,Pan2000,
Sanchez-Soto_2001,NeutronBell2003,Monroe2008,BellUFF2010,BellOAM,Leonardo2014,Stoklasa_2015,Balthazar2016,Quasar2018}. 
It is one of the pillars that distinguish quantum from classical correlations and also plays 
a major role in quantum information protocols. Since the seminal experiment by Aspect and co-workers \cite{Bell-Aspect}, many different 
tests of the Clauser-Horne-Shimmony-Holt (CHSH) inequality \cite{CHSH} have been performed with entangled photon pairs. 
These photonic tests of the CHSH criterion rely on intensity correlations measured with photodetectors, which are 
subject to the assumptions adopted in the realm of photodetection theory \cite{Loudon1973}. For example, intensity correlations are 
given by normally ordered correlation functions of the electromagnetic field operators. This normal ordering stems 
from the destructive nature of the detection mechanism through photon absorption. It also prevents any 
influence from the quantum vacuum, since its energy cannot be extracted by the photodetectors. In this sense, 
Bell's inequality violation can be affected by a vacuum sensitive detection system. 

Nevertheless, symmetric ordering is the prescription for constructing Hermitian operators composed by products of non 
commuting observables. In this context, phase-space quantum distributions play a key role in calculating averages, 
such as correlation functions, in a given operator ordering. For example, the Glauber P-distribution is associated 
with averages of normally ordered operator products, while the Wigner distribution corresponds to averages in 
symmetric ordering \cite{mandel_wolf_1995}.
Beyond a simple technical issue, this operator ordering has a more profound meaning regarding the vacuum contribution to 
intensity correlation measurements in quantum optical experiments. This is a long standing concern and we may quote an 
interesting discussion presented in Ref. \cite{Dalibard1982} about the role played by operator ordering in correctly 
accounting for the contribution of vacuum fluctuations in radiation reaction. More recently, operator ordering sensitivity 
has been related to the nonclassicality of bosonic field quantum states \cite{Kolobov2019}.

As we mentioned, photodetection signals naturally give 
normally ordered intensity correlations with vanishing vacuum contribution. Interestingly, the symmetrically ordered 
intensity correlations given by the Wigner representation encompass vacuum fluctuations and coincides with the 
results given by a classical stochastic model for the background noise \cite{Casado1998}. Such classical models have already 
been used to describe optical phenomena related to the vacuum fluctuations \cite{Casado1998,Olsen2001,Luciano2008,Casado_2019}.
Moreover, the quantum-classical boundary has been recently revisited in a number of research works with the aid of new 
theoretical tools for characterizing polarization in different optical degrees of freedom and the corresponding correlations 
\cite{Eberly2015,Eberly2016,Qian2016,Eberly2017,DeZela2018,Gonzales2018}. 

This paper aims to investigate the role played by operator ordering in Bell inequality violation with polarization 
entangled photons. We compare different ordering choices in the intensity correlations that figure in the CHSH 
inequality. We show that violation is precluded by symmetric ordering, which means that the \textit{blindness} of 
the photodectors to the quantum vacuum plays an important role.

\section{Quantum description of parametric down-conversion}

\begin{figure}
	\resizebox*{8.0cm}{5cm}{\includegraphics{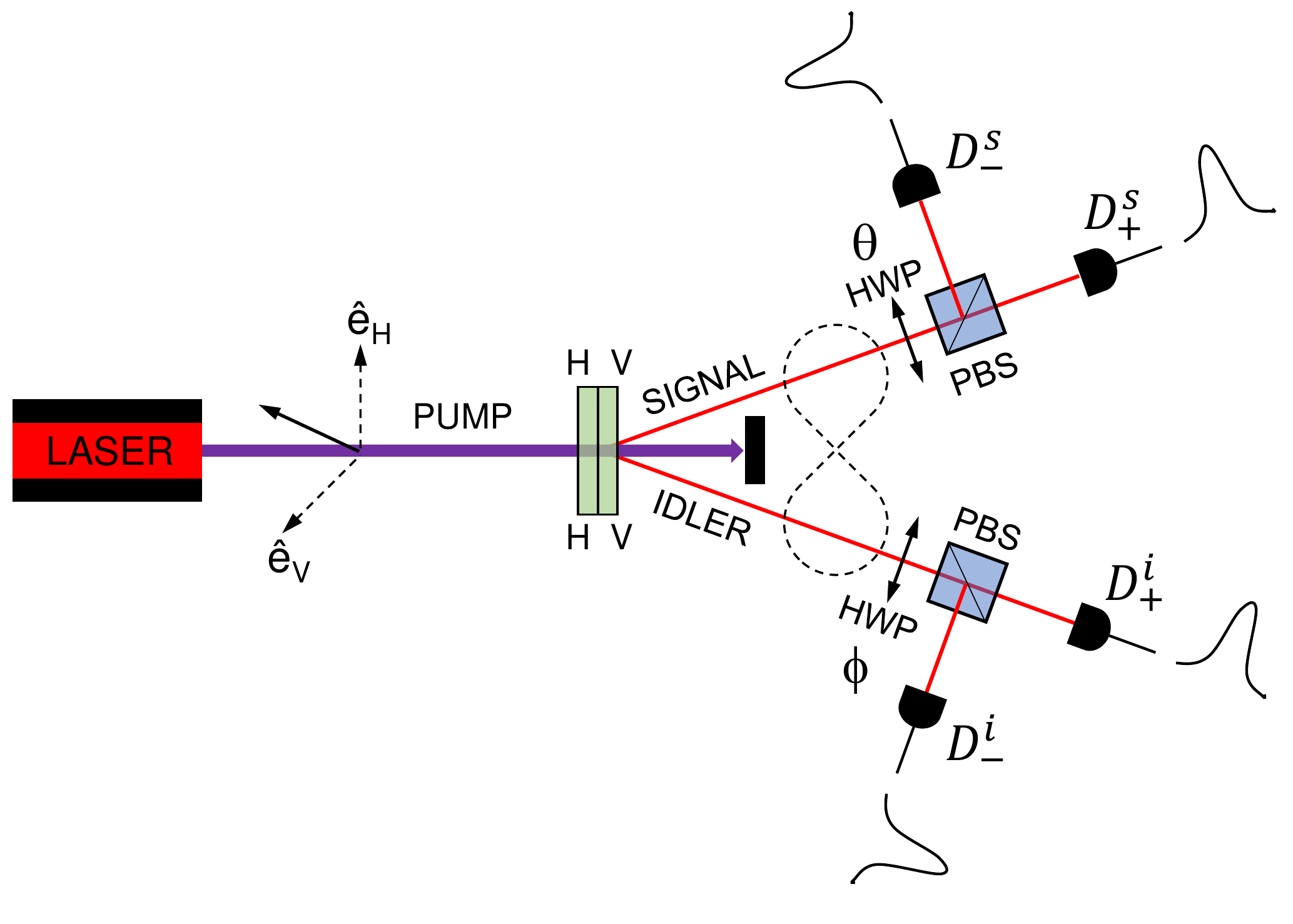}}
	\caption{Typical setup for Bell inequality measurements with a polarization entangled photon source. Two nonlinear crystals 
		are glued together 
	with their optical axes rotated with respect to each other. A pump beam polarized at $45^{o}$ can generate either horizontally 
	polarized photons in the first crystal or vertically polarized photons in the second. Two-photon polarization analysis is performed 
	in the detection region. Half-wave plates (HWP) are used to set the measurement angles $\theta$ and $\phi\,$. Polarization projection
	is performed with polarizing beam splitters (PBS) before the photons hit detectors $D_\pm^s$ and $D_\pm^i$. 
	When measuring symmetrically ordered intensity correlations, these detectors must be replaced by homodyne detection setups.}
	\label{fig:setup}
\end{figure}

Let us apply our ideas to a frequently used source of polarization entangled photon pairs \cite{Kwiat1995,Santos2001,Juliana2003}. 
It is composed 
by two identical nonlinear crystals glued together with their optical axes rotated by $90^o$ relative to each other. 
A strong laser beam at frequency $\omega$ is used to pump the crystals and generate photon pairs at frequency $\omega/2$ by 
spontaneous parametric down conversion (SPDC). The pump beam is 
linearly polarized at $45^o$ with respect to the crystals axes. Under type-I phase match, a pair of linearly polarized 
photons is generated either with horizontal polarization in one crystal or with vertical polarization in the other, thus 
producing a polarization entangled state. Two polarizers are used in the detection region, one before each photodetector, 
to set the measurement bases. This setup is depicted in Fig. \ref{fig:setup}. This will be our model system for investigating 
the CHSH inequality under different correlation ordering and the role played by quantum vacuum. For symmetrically ordered 
intensity correlations, detectors $D_\pm^s$ and $D_\pm^i$ must be replaced by homodyne detection setups, as will be explained 
in Section \ref{measurement}.

For a thin crystal and low nonlinear susceptibility $\chi$, the pump laser is very little affected by the down-conversion process and only a small 
fraction of the incoming photons is converted into photon pairs. In this case, we can assume that the input quantum state of the 
pump beam remains unaltered by the parametric interaction. Moreover, we will consider the pump laser polarized at $45^o\,$, prepared in a coherent 
state $\ket{v_p}_H\otimes\ket{v_p}_V\,$. The Hamiltonians for the SPDC process in crystals 1 and 2 are,
\begin{eqnarray}
  \hat{H_1} = i \hbar g\, \hat{a}^{s\,\dagger}_{H}\hat{a}^{i\,\dagger}_{H} + \mathrm{h.c.} \,,
\nonumber\\
  \hat{H_2} = i \hbar g\, \hat{a}^{s\,\dagger}_{V}\hat{a}^{i\,\dagger}_{V} + \mathrm{h.c.} \,,
\label{hamiltoniana}
\end{eqnarray}
where $g=\chi v_p\,$ is the nonlinear coupling constant, $\hat{a}^s_j$, $\hat{a}^i_j$ ($j=H,V$) are 
boson operators for signal and idler modes with horizontal ($H$) and vertical ($V$) polarizations, 
and h.c. stands for Hermitian conjugate.

In either Glauber or Wigner representation, the density matrix $\hat{\rho}$ is represented by a quasiprobability 
distribution $P(\mathbf{a}, \mathbf{a}^*)$ (Glauber) or $W(\mathbf{a}, \mathbf{a}^*)$ (Wigner) for a column vector of 
complex stochastic amplitudes $\mathbf{a}=(a^s_H,a^s_V,a^i_H,a^i_V)^T\,$. These distributions are 
readily obtained from the density matrix through the corresponding characteristic function. The Glauber representation 
is given as the Fourier transform of the normally ordered characteristic function $C_P$ \cite{mandel_wolf_1995}
\begin{eqnarray}
P(\mathbf{a},\mathbf{a}^*) &=& \frac{1}{\pi^2}\int C_P (\mathbf{z},\mathbf{z}^*)\,
e^{-i\left(\mathbf{z}^{*}\mathbf{a}^* + \mathbf{z}\,\mathbf{a}\right)} d^2\mathbf{z}\;,
\nonumber\\
C_P(\mathbf{z},\mathbf{z}^*) &=& \mathrm{Tr}\left[\hat{\rho}\,e^{i\left(\mathbf{z}^{*}\mathbf{\hat{a}}^{\dagger}\right)}
\,e^{i\left(\mathbf{z}\,\mathbf{\hat{a}}\right)}\right]\;,
\end{eqnarray}
where $\mathbf{z}=(z^s_H,z^s_V,z^i_H,z^i_V)$ is a row vector of Fourier variables, one for each mode amplitude, 
and $\mathbf{\hat{a}}=(\hat{a}^s_H,\hat{a}^s_V,\hat{a}^i_H,\hat{a}^i_V)^T$ is a column vector with the 
corresponding annihilation operators. Averages of normally ordered operator products for any mode $j$ are readily calculated 
with the Glauber distribution as follows
\begin{eqnarray}
\langle \hat{a}_j^{\dagger\,m}\,\hat{a}_j^{n}\rangle_N &=& \int P(\mathbf{a},\mathbf{a}^*)\,\, a_j^{*\,m} a_j^n\,\,d^2\mathbf{a}\;.
\end{eqnarray}

The Wigner representation 
is given as the Fourier transform of the symmetrically ordered characteristic function $C_W$ \cite{mandel_wolf_1995}
\begin{eqnarray}
W(\mathbf{a},\mathbf{a}^*) &=& \frac{1}{\pi^2}\int C_W (\mathbf{z},\mathbf{z}^*)\,
e^{-i\left(\mathbf{z}^{*}\mathbf{a}^* + \mathbf{z}\,\mathbf{a}\right)} d^2\mathbf{z}\;,
\nonumber\\
C_W(\mathbf{z},\mathbf{z}^*) &=& \mathrm{Tr}\left[\hat{\rho}\,
e^{i\left(\mathbf{z}^{*}\mathbf{\hat{a}}^{\dagger}+\mathbf{z}\,\mathbf{\hat{a}}\right)}\right]\;.
\end{eqnarray}
Averages of symmetrically ordered operator products for any mode $j$ are readily calculated 
with the Wigner distribution as follows
\begin{eqnarray}
\langle \hat{a}_j^{\dagger\,m}\,\hat{a}_j^{n}\rangle_S &=& \int W(\mathbf{a},\mathbf{a}^*)\,\, a_j^{*\,m} a_j^n\,\,d^2\mathbf{a}\;.
\end{eqnarray}

For the Hamiltonians given in Eqs. \eqref{hamiltoniana}, the dynamics of SPDC can be described by a 
Fokker-Planck equation for the time evolution of both Glauber and Wigner distributions 
\cite{carmichael_1999,Dodonov_2003,WignerUFF2010}. 
From the corresponding Fokker-Planck equation for $P$ or $W\,$, one can derive a set of Langevin 
equations for the stochastic amplitudes $\mathbf{a}\,$. Neglecting losses in the thin crystal regime, these equations 
are
\begin{eqnarray}
\dot{a}^s_H &=& g a^{i\,*}_H\,,\qquad \dot{a}^s_V = g a^{i\,*}_V\;,
\nonumber\\
\dot{a}^i_H &=& g a^{s\,*}_H\,,\qquad \dot{a}^i_V = g a^{s\,*}_V\;,
\label{eq.estocast2}
\end{eqnarray}
where the equations on the left are for crystal 1 and those on the right are for crystal 2.
The main difference between the two representations consists in the correlation functions between the input 
amplitudes, as we will make explicit shortly.

The pump and down converted electric fields will be described by plane waves of the form
\begin{eqnarray}
\mathbf{E}_p(\mathbf{r},t) &=& v_p\left(\hat{\mathbf{e}}_H + \hat{\mathbf{e}}_V\right)\, 
e^{i(\textbf{k}_{p} \cdot \textbf{r} - \omega_{p} t)} \;,
\nonumber
\\
\mathbf{E}_s(\mathbf{r},t) &=& \left(a^s_H \hat{\mathbf{e}}_H + a^s_V \hat{\mathbf{e}}_V\right)\, 
e^{i(\textbf{k}_{s} \cdot \textbf{r} - \omega_{s} t)} \;,
\nonumber
\\
\mathbf{E}_i(\mathbf{r},t) &=& \left(a^i_H \hat{\mathbf{e}}_H + a^i_V \hat{\mathbf{e}}_V\right)\, 
e^{i(\textbf{k}_{i} \cdot \textbf{r} - \omega_{i} t)} \;,
\label{1}
\end{eqnarray}
where $\mathbf{k}_j$ ($j=p,s,i$) is the wave vector of the corresponding mode. We will assume perfect 
phase match between pump and down converted fields, so that $\mathbf{k}_p = \mathbf{k}_s + \mathbf{k}_i\,$. 
The input downconverted fields, signal and idler, are assumed to be in vacuum state and the initial values 
of their amplitudes are complex, Gaussian-distributed stochastic variables that simulate the incoming vacuum 
fluctuations. These amplitudes obey the following correlations:
\begin{eqnarray}
\langle a^j_{0k} \rangle &=& 0\;,
\nonumber\\ 
\langle a^j_{0k} a^{j^\prime}_{0k} \rangle &=& 0\;,
\nonumber\\
\langle a^j_{0k} a^{j^\prime*}_{0k^\prime} \rangle &=& \frac{\epsilon}{2} \delta_{jj^\prime}\delta_{kk^\prime}\;, 
\label{5}
 \end{eqnarray}
where $j,j^\prime = s,i\,$; $k,k^\prime = H,V$ and $\epsilon = 0$ $(1)$ for Glauber (Wigner) representation.

We next solve the dynamical equations for the signal and idler complex amplitudes after passage through both 
crystals.

\subsection{Fisrt crystal}

The first crystal converts a vertically polarized pump photon into a pair of horizontally polarized signal and idler photons. 
The interaction time is $\tau=nd/c\,$, where $d$ is the propagation distance inside the crystal, $n$ is the refractive index 
and $c$ is the speed of light in vacuum. The output complex amplitudes are given by the solution of the left 
Eqs.\eqref{eq.estocast2}:
\begin{eqnarray}
\nonumber
a^s_{H}(\tau) &=& a^s_{0H} \cosh{g \tau} + a^{i*}_{0H} \sinh{g \tau}, \\
a^i_{H}(\tau) &=& a^{i}_{0H} \cosh{g \tau} + a^{s*}_{0H} \sinh{g \tau}.
\label{3.3}
\end{eqnarray}
Note that after passing the first crystal, the signal and idler amplitudes $a^s_H$ and $a^i_H$ exhibit a cross-talk between 
their incoming vacuum fluctuations. This is crucial for understanding the origin of the correlations (entanglement) between 
signal and idler as a vacuum induced effect.
Using the input correlations given in Eq. \ref{5}, the output correlations after interaction in the first crystal are
\begin{eqnarray}
\langle a^j_{H}(\tau) \rangle &=& 0 \;,
\nonumber\\  
\langle a^{j\,*}_{H}(\tau) a^{j\prime}_{H}(\tau) \rangle &=& \left(\frac{\epsilon}{2} + \sinh^2{g \tau}\right)\,\delta_{jj^\prime}\;,
\nonumber\\ 
\langle a^s_{H}(\tau) a^{i}_{H}(\tau) \rangle &=& \cosh{g \tau} \sinh{g \tau},
\label{3.4}
\end{eqnarray}
where $j,j^\prime = s,i\,$.

\subsection{Second crystal}

The second crystal converts a horizontally polarized pump photon into a pair of vertically polarized signal and idler photons. 
The interaction time is also $\tau$ if we assume the crystals have the same width. 
The output complex amplitudes are given by the solution of the right Eqs.\eqref{eq.estocast2}:
\begin{eqnarray}
\nonumber
a^s_{V}(\tau) &=& a^s_{0V} \cosh{g \tau} + a^{i*}_{0V} \sinh{g \tau}, \\
a^i_{V}(\tau) &=& a^{i}_{0V} \cosh{g \tau} + a^{s*}_{0V} \sinh{g \tau}.
\label{4.3}
\end{eqnarray}
After passing the second crystal, the signal and idler amplitudes $a^s_V$ and $a^i_V$ also exhibit a cross-talk between 
their incoming vacuum fluctuations, inducing correlations (entanglement).
Using the input correlations given in Eq. \eqref{5}, the output correlations after interaction in the second crystal are
\begin{eqnarray}
\langle a^j_{V}(\tau) \rangle &=& 0 \;,
\nonumber\\  
\langle a^{j\,*}_{V}(\tau) a^{j\prime}_{V}(\tau) \rangle &=& \left(\frac{\epsilon}{2} + \sinh^2{g \tau}\right)\,\delta_{jj^\prime}\;,
\nonumber\\ 
\langle a^s_{V}(\tau) a^{i}_{V}(\tau) \rangle &=& \cosh{g \tau} \sinh{g \tau},
\label{4.4}
\end{eqnarray}
where $j,j^\prime = s,i\,$.

\subsection{Polarization measurement settings}

After leaving the crystals, the entangled photons travel to the detectors region and traverse two polarizing beam splitters (PBS) 
preceded by half-wave plates that set the measurement angles at $\theta$ (signal) and $\phi$ (idler) before hitting the detectors. 
After passing the respective HWP and PBS, each beam will be divided into two polarization components that mix the input $H$ and 
$V$ amplitudes, producing the following rotated variables
\begin{eqnarray}
a^s_+(\theta) &=& a^s_{H} \cos{\theta} + a^s_{V} \sin{\theta}\;, 
\nonumber\\
a^s_-(\theta ) &=& a^s_{V} \cos{\theta} - a^s_{H} \sin{\theta}\;,
\label{5.1}\\
a^i_+(\phi) &=& a^i_{H} \cos{\phi} + a^i_{V}  \sin{\phi}\;,
\nonumber\\
a^i_-(\phi) &=& a^i_{V} \cos{\phi} - a^i_{H}  \sin{\phi}\;.
\label{5.2}
\end{eqnarray}
The electric field at detectors $D^s_{\pm}$ and $D^i_{\pm}$ will be given by
\begin{eqnarray}
E_s (\textbf{r}^s_+,t) &=& a^s_+(\theta)\,e^{i\left(\textbf{k}_{s} \cdot \textbf{r}^s_+ - \omega_s t\right)}\;,
\nonumber\\
E_s (\textbf{r}^s_-,t) &=& a^s_-(\theta)\,e^{i\left(\textbf{k}_{s} \cdot \textbf{r}^s_- - \omega_s t\right)}\;,
\nonumber\\
E_i(\textbf{r}^i_+,t) &=& a^i_+(\phi)\,e^{i\left(\textbf{k}_{i} \cdot \textbf{r}^i_+ - \omega_i t\right)}\;,
\nonumber\\
E_i(\textbf{r}^i_-,t) &=& a^i_-(\phi)\,e^{i\left(\textbf{k}_{i} \cdot \textbf{r}^i_- - \omega_i t\right)}\;.
\label{5.3}
\end{eqnarray}
These rotated amplitudes will determine the polarization correlations that figure in the CHSH inequality.

\section{Field correlations}

The CHSH criterion for Bell violation is evaluated from coincidence measurements that correspond to intensity correlations 
between signal and idler. We now calculate several correlation functions in the Glauber and Wigner representations.

\subsection{Individual Intensities}

First, we calculate the field intensity at each detector
\begin{eqnarray}
  \langle I^s_\pm(\theta) \rangle &=& \langle E^*_s(\textbf{r}^s_\pm)E_s(\textbf{r}^s_\pm)
  \rangle = \langle a^{s\,*}_\pm(\theta)a^s_\pm(\theta) \rangle \;,
  \nonumber \\
  \langle I^i_\pm(\phi) \rangle &=& \langle E^*_i(\textbf{r}^i_\pm)E_i(\textbf{r}^i_\pm)
  \rangle = \langle a^{i\,*}_\pm(\phi)a^i_\pm(\phi) \rangle \;,
\label{6.1}
\end{eqnarray}
Substituting expressions Eqs. \eqref{5.1} and \eqref{5.2}, and using the amplitudes correlations given by Eqs. \eqref{3.4} e \eqref{4.3},
we arrive at
\begin{eqnarray}
\langle I^s_\pm(\theta)\rangle = \langle I^i_\pm(\phi) \rangle = \frac{\epsilon}{2} + \sinh^2{g\tau}\;. 
\label{6.2}
\end{eqnarray}
Note that the individual intensities are insensitive to the polarizers settings. This will be different for the intensity 
correlations as we show next.

\subsection{Intensity correlations}

The input quantum fluctuations on signal and idler fields are uncorrelated before entering the crystals. However, after undergoing parametric 
interaction within the crystals, their amplitudes become correlated according to Eqs. \eqref {3.4} and \eqref {4.3}. 
The intensity correlations measured on two separate detectors are given by
\begin{eqnarray}
C_{jk}(\theta, \phi) &=&   \langle E^*_s(\textbf{r}^s_j)E_s(\textbf{r}^s_j)
E^*_i(\textbf{r}^i_k)E_i(\textbf{r}^i_k)  \rangle
\nonumber\\
&=& 
\langle  I^s_j(\theta) I^i_k(\phi) \rangle\;,
\label{6.3}
\end{eqnarray}
where  $ j = \pm $ and $ k = \pm\,$. These intensity correlations can be calculated by using 
the following relationship that holds for stochastic Gaussian variables
\begin{eqnarray}
\langle a_1 a_2 a_3 a_4 \rangle &=& \langle a_1a_2 \rangle \langle a_3 a_4 \rangle +
\langle a_1a_3\rangle \langle a_2a_4 \rangle 
\nonumber\\
&+& \langle a_1a_4 \rangle \langle a_2a_3 \rangle\;.
\label{6.4} 
\end{eqnarray}
With the aid of relation \eqref{6.4}, the two-photon polarization correlations can be written as 
\begin{eqnarray}
C_{jk}(\theta, \phi) &=&  \langle a^{s\,*}_j(\theta)a^s_j(\theta) \rangle \langle a^{i\,*}_k(\phi)a^i_k(\phi) \rangle
\nonumber\\
&+&  \langle a^{s\,*}_j(\theta)a^{i\,*}_k(\phi)\rangle \langle a^s_j(\theta)a^i_k(\phi) \rangle
\nonumber\\
&+& \langle a^{s\,*}_j(\theta)a^i_k(\phi) \rangle \langle a^s_j(\theta)a^{i\,*}_k(\phi) \rangle \;.
\label{6.7}
\end{eqnarray}
Using now the correlations given by Eqs. \eqref{3.4} and \eqref{4.4}, we find
\begin{eqnarray}
C_{++}(\theta, \phi) &=& C_{--}(\theta, \phi) 
\label{6.8}
\\
&=& \frac{\left(\frac{\epsilon}{2} + \sinh^2g\tau\right)^2 + \sinh^2{2g\tau}\,\cos^2{(\theta - \phi)}}{4}\;,
\nonumber
\\
C_{+-}(\theta, \phi) &=& C_{-+}(\theta, \phi) 
\label{6.9}
\\
&=& \frac{\left(\frac{\epsilon}{2} + \sinh^2g\tau\right)^2 + \sinh^2{2g\tau}\,\sin^2{(\theta - \phi)}}{4}\;.
\nonumber
\end{eqnarray}

We next evaluate the impact of operator ordering on the CHSH criterion for the quantum-classical correlation boundary.

\section{Bell inequality}

Let us apply the Clauser-Horne-Shimony-Holt (CHSH) inequality to the two-photon polarization correlations 
and compare the results obtained with the symmetric and normal operator ordering. The correlations obtained 
at a given measurement setting with angles $\theta$ (signal) and $\phi$ (idler) are given by
\begin{eqnarray}
M (\theta, \phi) = \frac{C_{++}  + C_{--} - C_{+-} - C_{-+}}{C_{++}  + C_{--} + C_{+-} + C_{-+}}\;.
\label{7.1}
\end{eqnarray}
Then, the CHSH inequality is evaluated for the quantity
\begin{equation}
S = M (\theta, \phi)  + M (\theta^\prime, \phi)  - M (\theta, \phi^\prime)  + M (\theta^\prime, \phi^\prime)\;. 
\label{7.2}
\end{equation}
Classical correlations are restricted to $-2 \leq S \leq 2$. However, this inequality can be violated 
for quantum correlated polarization modes, where maximum violation occurs when $S = 2\sqrt{2}\,$. This can be 
accomplished with the following polarization settings:
$\theta = 0$, $\theta^\prime = \pi/4$, $\phi = \pi/8$ e $\phi^\prime= 3 \pi/8\,$, for example.
We next check the CHSH inequality in each operator ordering by plugging \eqref{6.8} and \eqref{6.9} 
into \eqref{7.1} and \eqref{7.2}. Note that the ordering dependent terms in Eqs. \eqref{6.8} and \eqref{6.9} 
cancel out in the numerator of \eqref{7.1} but they do contribute to the denominator. As we show below, 
it drastically affects the violation of the CHSH criterion in the thin crystal limit $g\tau\ll 1\,$, 
usually valid in actual experimental conditions. 

\begin{figure}
    \includegraphics[width=8cm]{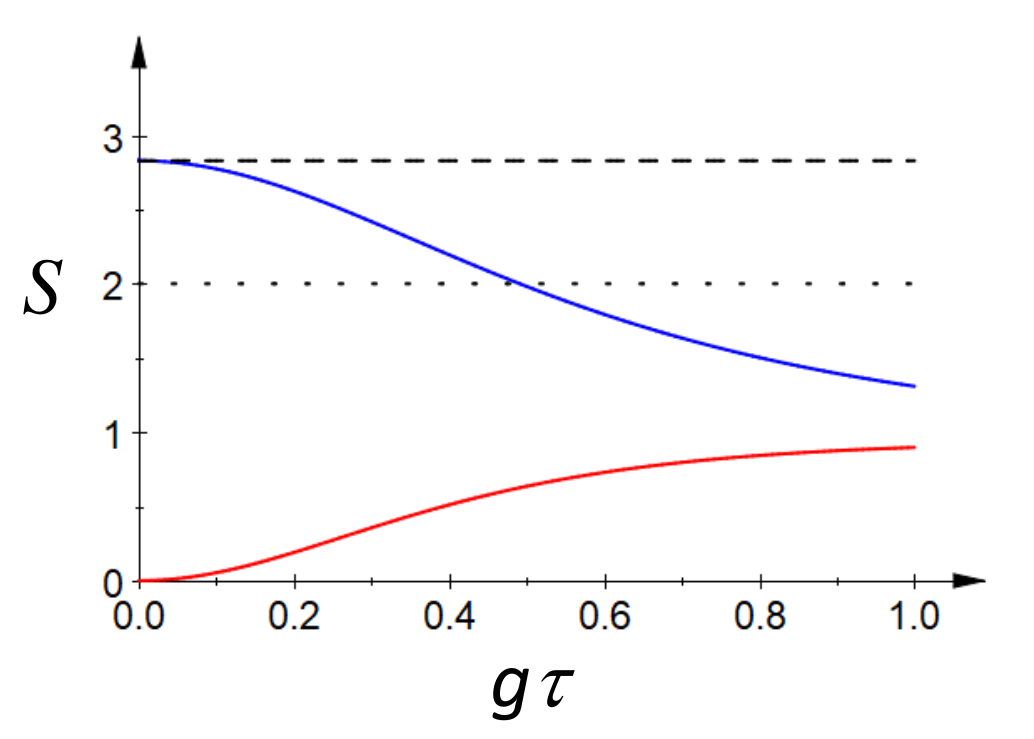}
	\caption{Bell parameter $S$ for normal (blue online) and symmetric (red online) ordering. 
		The dashed and dotted lines show the $2\sqrt{2}$ limit and 
	the violation boundary, respectively.}
	\label{fig:grafico}
\end{figure}

\begin{itemize}
	
\item Normal ordering
\begin{eqnarray}
&& M (\theta, \phi) = \left(\frac{\sinh^2 2g\tau}{\sinh^2 2g\tau + 8\sinh^4 g\tau}\right)\,\cos{[2(\theta - \phi)]}\;,
\nonumber\\
&& S_N = 2\sqrt{2}\,\left(\frac{\sinh^2 2g\tau}{\sinh^2 2g\tau + 8\sinh^4 g\tau}\right)\approx 2\sqrt{2} \,.
\label{7.3}
\end{eqnarray}
\vspace{0.1cm}

\item Symmetric ordering
\begin{eqnarray}
&& M (\theta, \phi) = \left(\frac{\sinh^2 2g\tau}{\sinh^2 2g\tau + 2\cosh^2 2g\tau}\right)\,\cos{[2(\theta - \phi)]}\;,
\nonumber\\
&& S_S = 2\sqrt{2}\,\left(\frac{\sinh^2 2g\tau}{\sinh^2 2g\tau + 2\cosh^2 2g\tau}\right)\approx 0\,.
\label{7.4}
\end{eqnarray}
\end{itemize}
As we can see, in the thin crystal limit ($g\tau\ll 1$) the normally ordered polarization correlations give maximal 
violation of the CHSH inequality, while the symmetric ordered polarization correlations do not violate. This can be 
easily visualized in the graphic shown in Fig. \ref{fig:grafico}, where the CHSH quantity $S$ given by Eqs. \eqref{7.3} 
and \eqref{7.4} is plotted as a function of $g\tau\,$. 

\section{Measurement Scheme}
\label{measurement}

A natural question to be asked is whether the intensity correlations between the vacuum modes predicted in the 
symmetric order can be accessed by some mechanism. 
Let us recall that symmetric ordering is the quantum theory prescription when calculating averages of physical 
quantities composed by functions of non-commuting observables. This is the case, for example, of the intensity 
\begin{eqnarray}
I = X^2 + Y^2 = \frac{a a^\dagger + a^\dagger a}{2}\;,
\end{eqnarray}
where we defined the quadratures
\begin{eqnarray}
X = \frac{a + a^\dagger}{2}\,, \qquad\qquad Y = \frac{a - a^\dagger}{2i}\,.
\end{eqnarray}
This quantity is not the one measured in the standard photodetection scheme, which, being based on photon absorption, 
is not sensitive to anti-normal terms. Therefore, the usual schemes have limited access to the field fluctuations, 
since they are blind to the vacuum field contribution.
In order to obtain full information about the field properties by measuring the symmetrically ordered 
correlation functions, one must perform a direct measurement of the field amplitude, 
more precisely the field quadratures. 
In the radio frequency range of the electromagnetic spectrum, the response of a regular antenna is indeed proportional 
to the field amplitude and the quadratures can be directly measured. 
However, discretization of the energy exchange in this regime is negligible and quantum effects 
cannot be sensed. Meanwhile, direct detection of the fast field oscillations in the optical regime is challenging 
and one must resort to homodyne measurements. 
An interesting discussion on experimental techniques for measuring 
correlation functions in normal, symmetric and anti-normal ordering is presented in Refs. \cite{LesHouches2013,Stiller2014}, 
where a direct correspondence between operator ordering and detection schemes is summarized as follows
\begin{itemize}
	\item \textbf{normal ordering} $a^\dagger a$: direct detection.
	\item \textbf{symmetric ordering} $(a^\dagger a + a a^\dagger)/2$: homodyne detection.
	\item \textbf{anti-normal ordering} $a a^\dagger$: heterodyne detection.
\end{itemize}
In the heterodyne (double-homodyne) detection scheme, it is possible to simultaneously measure canonical conjugate 
quadratures, at the expense of allowing extra (vacuum) noise into the detection mechanism.

\begin{figure}
	\includegraphics[width=6cm]{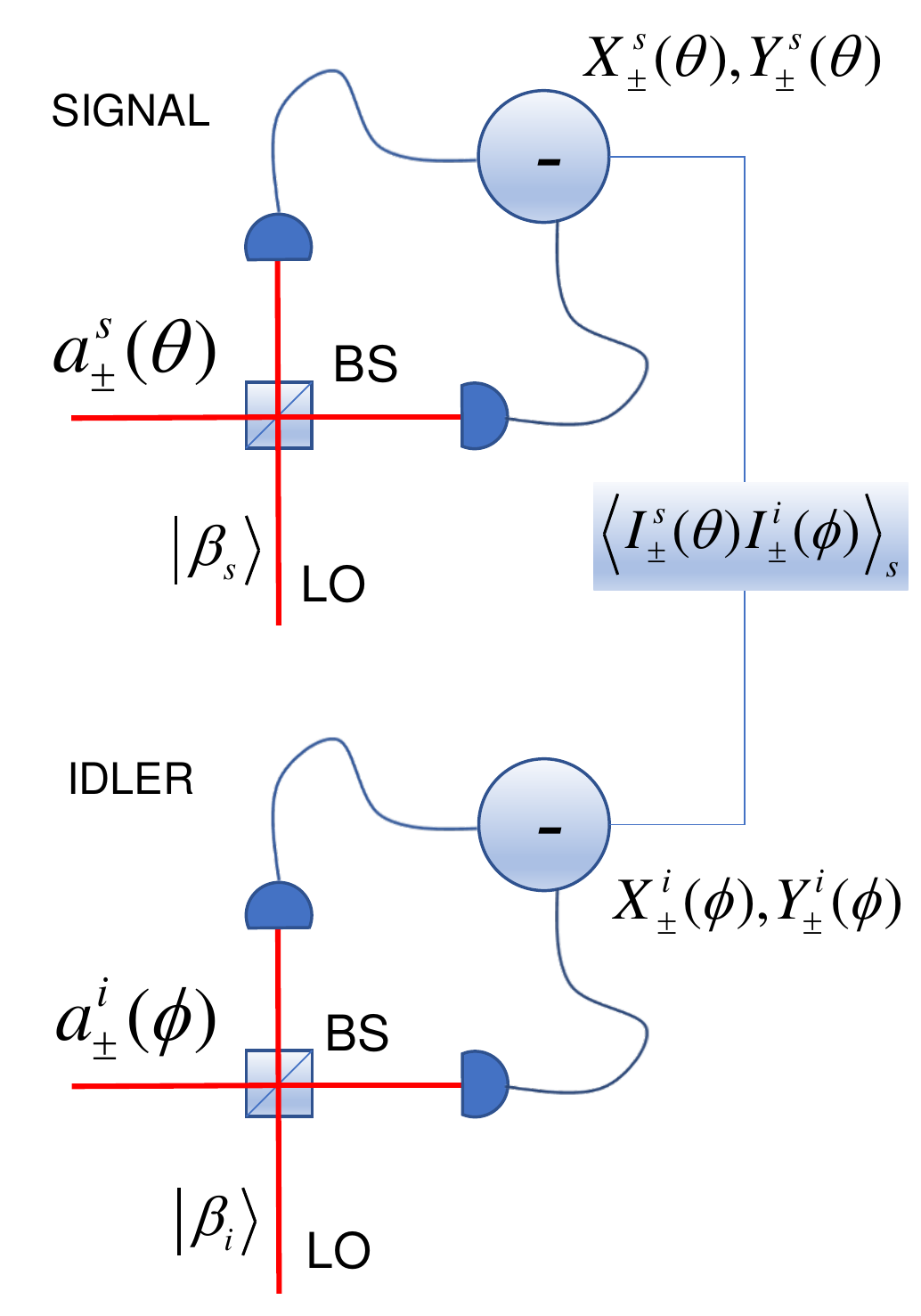}
	\caption{Measurement scheme for symmetrically ordered correlations. BS: Beam splitter, LO: Local oscillator.}
	\label{fig:measurement}
\end{figure}

The symmetrically ordered averages can be accessed with the homodyne measurement setup depicted in Fig. \ref{fig:measurement}. 
Each polarization output of signal and idler is sent to a homodyne detection setup, where it is mixed with a local oscillator 
prepared in a coherent state $\ket{\beta_j}$ ($j=s,i$). By adjusting the phase of each local oscillator independently, all 
quadrature combinations $(X^s_\pm,X^i_\pm)$, $(X^s_\pm,Y^i_\pm)$, $(Y^s_\pm,X^i_\pm)$, $(Y^s_\pm,Y^i_\pm)$, 
can be measured for each setting of the polarization analyzers. Then, the homodyne detection 
data can be processed to compute the symmetrically ordered intensity correlations from
\begin{eqnarray}
\!\!\!\moy{I_{\pm}^s(\theta) I_{\pm}^i(\phi)}_{s} \!=\! 
\left\langle\left[(X_\pm^{s})^2 \!+\! (Y_\pm^s)^2\right]\!\!\left[(X_\pm^i)^2 \!+\! (Y_\pm^i)^2\right]\right\rangle .
%\nonumber\\
\end{eqnarray}
A recent measurement of the CHSH inequality for continuous variables employed a homodyne detection scheme similar 
to the one described here \cite{PingKoyLam2018}. 

\section{conclusion}
\label{conclusion}

In conclusion, we analyzed Bell's inequality violation in two-photon polarization correlations under different 
operator ordering of the intensity correlation functions.  
Under the usual experimental condition of weak parametric coupling (thin crystal), the 
normally ordered intensity correlations violate the CHSH criterion, while the symmetrically ordered ones do not. Beyond a 
technical issue, this operator ordering has a more profound physical meaning.

Normal ordering is imposed by photodetection signals based on photon absorption and therefore precludes any influence from 
the quantum vacuum, which energy cannot be extracted. Meanwhile, symmetric ordering is the usual prescription for 
constructing Hermitian operators from products of non commuting observables. Moreover, symmetric ordering is crucial for 
evidencing vacuum effects like Casimir force, spontaneous emission, among many others \cite{Milonni1994}. 
The symmetrically ordered intensity correlations can be measured with the detection schemes discussed in 
Refs. \cite{LesHouches2013,Stiller2014}. We can also quote a recent measurement of the vacuum fluctuations,
as reported in Ref.\cite{Vacuum-measure}. 

As a final remark, it is worthwhile to mention that for sufficiently large parametric interaction ($g\tau\sim 1$), 
no CHSH violation is predicted in either operator ordering. However, this regime falls outside the validity of the 
non depletion assumption for the pump beam. In this case, the triple interaction between pump, signal and idler must 
be fully solved, giving rise to triple correlations that may generate tripartite entanglement witnessed by other 
inequality criteria \cite{Cassemiro:07,Coelho2009}. 
For example, strong parametric interaction is attained in four-wave mixing sources of entangled photon 
pairs \cite{Lett2008,Boyer2008,Ferraz2005}. 
The discussion draw in this article will be pursued to the strong interaction regime in future contributions.

\section*{Acknowledgments}
Funding was provided by 
Conselho Nacional de Desenvolvimento Cient\'{\i}fico e Tecnol\'ogico (CNPq), 
Coordena\c c\~{a}o de Aperfei\c coamento de 
Pessoal de N\'\i vel Superior (CAPES), Funda\c c\~{a}o Carlos Chagas Filho de Amparo \`{a} 
Pesquisa do Estado do Rio de Janeiro (FAPERJ), and Instituto Nacional 
de Ci\^encia e Tecnologia de Informa\c c\~ao Qu\^antica (INCT-IQ).

\bibliography{bell-vacuum-3}

\begin{thebibliography}{47}
\expandafter\ifx\csname natexlab\endcsname\relax\def\natexlab#1{#1}\fi
\expandafter\ifx\csname bibnamefont\endcsname\relax
  \def\bibnamefont#1{#1}\fi
\expandafter\ifx\csname bibfnamefont\endcsname\relax
  \def\bibfnamefont#1{#1}\fi
\expandafter\ifx\csname citenamefont\endcsname\relax
  \def\citenamefont#1{#1}\fi
\expandafter\ifx\csname url\endcsname\relax
  \def\url#1{\texttt{#1}}\fi
\expandafter\ifx\csname urlprefix\endcsname\relax\def\urlprefix{URL }\fi
\providecommand{\bibinfo}[2]{#2}
\providecommand{\eprint}[2][]{\url{#2}}

\bibitem[{\citenamefont{Freedman and Clauser}(1972)}]{Freedman1972}
\bibinfo{author}{\bibfnamefont{S.~J.} \bibnamefont{Freedman}} \bibnamefont{and}
  \bibinfo{author}{\bibfnamefont{J.~F.} \bibnamefont{Clauser}},
  \bibinfo{journal}{Phys. Rev. Lett.} \textbf{\bibinfo{volume}{28}},
  \bibinfo{pages}{938} (\bibinfo{year}{1972}),
  \urlprefix\url{https://link.aps.org/doi/10.1103/PhysRevLett.28.938}.

\bibitem[{\citenamefont{Aspect et~al.}(1981)\citenamefont{Aspect, Grangier, and
  Roger}}]{Bell-Aspect}
\bibinfo{author}{\bibfnamefont{A.}~\bibnamefont{Aspect}},
  \bibinfo{author}{\bibfnamefont{P.}~\bibnamefont{Grangier}}, \bibnamefont{and}
  \bibinfo{author}{\bibfnamefont{G.}~\bibnamefont{Roger}},
  \bibinfo{journal}{Phys. Rev. Lett.} \textbf{\bibinfo{volume}{47}},
  \bibinfo{pages}{460} (\bibinfo{year}{1981}),
  \urlprefix\url{https://link.aps.org/doi/10.1103/PhysRevLett.47.460}.

\bibitem[{\citenamefont{Bell and Aspect}(2004)}]{bell_aspect_2004}
\bibinfo{author}{\bibfnamefont{J.~S.} \bibnamefont{Bell}} \bibnamefont{and}
  \bibinfo{author}{\bibfnamefont{A.}~\bibnamefont{Aspect}},
  \emph{\bibinfo{title}{Speakable and Unspeakable in Quantum Mechanics:
  Collected Papers on Quantum Philosophy}} (\bibinfo{publisher}{Cambridge
  University Press}, \bibinfo{year}{2004}), \bibinfo{edition}{2nd} ed.

\bibitem[{\citenamefont{Kwiat et~al.}(1993)\citenamefont{Kwiat, Steinberg, and
  Chiao}}]{Kwiat-Steinberg_1993}
\bibinfo{author}{\bibfnamefont{P.~G.} \bibnamefont{Kwiat}},
  \bibinfo{author}{\bibfnamefont{A.~M.} \bibnamefont{Steinberg}},
  \bibnamefont{and} \bibinfo{author}{\bibfnamefont{R.~Y.} \bibnamefont{Chiao}},
  \bibinfo{journal}{Physical review. A} \textbf{\bibinfo{volume}{47}},
  \bibinfo{pages}{R2472} (\bibinfo{year}{1993}).

\bibitem[{\citenamefont{Kwiat et~al.}(1994)\citenamefont{Kwiat, Eberhard,
  Steinberg, and Chiao}}]{Kwiat-Steinberg_1994}
\bibinfo{author}{\bibfnamefont{P.~G.} \bibnamefont{Kwiat}},
  \bibinfo{author}{\bibfnamefont{P.~H.} \bibnamefont{Eberhard}},
  \bibinfo{author}{\bibfnamefont{A.~M.} \bibnamefont{Steinberg}},
  \bibnamefont{and} \bibinfo{author}{\bibfnamefont{R.~Y.} \bibnamefont{Chiao}},
  \bibinfo{journal}{Physical review. A} \textbf{\bibinfo{volume}{49}},
  \bibinfo{pages}{3209} (\bibinfo{year}{1994}).

\bibitem[{\citenamefont{Weihs et~al.}(1998)\citenamefont{Weihs, Jennewein,
  Simon, Weinfurter, and Zeilinger}}]{Zeilinger1998}
\bibinfo{author}{\bibfnamefont{G.}~\bibnamefont{Weihs}},
  \bibinfo{author}{\bibfnamefont{T.}~\bibnamefont{Jennewein}},
  \bibinfo{author}{\bibfnamefont{C.}~\bibnamefont{Simon}},
  \bibinfo{author}{\bibfnamefont{H.}~\bibnamefont{Weinfurter}},
  \bibnamefont{and}
  \bibinfo{author}{\bibfnamefont{A.}~\bibnamefont{Zeilinger}},
  \bibinfo{journal}{Phys. Rev. Lett.} \textbf{\bibinfo{volume}{81}},
  \bibinfo{pages}{5039} (\bibinfo{year}{1998}),
  \urlprefix\url{https://link.aps.org/doi/10.1103/PhysRevLett.81.5039}.

\bibitem[{\citenamefont{J.~W.~Pan and Zeilinger}(2000)}]{Pan2000}
\bibinfo{author}{\bibfnamefont{M.~D. H.~W.} \bibnamefont{J.~W.~Pan},
  \bibfnamefont{D.~Bouwmeester}} \bibnamefont{and}
  \bibinfo{author}{\bibfnamefont{A.}~\bibnamefont{Zeilinger}},
  \bibinfo{journal}{Nature} \textbf{\bibinfo{volume}{403}},
  \bibinfo{pages}{515} (\bibinfo{year}{2000}),
  \urlprefix\url{https://doi.org/10.1038/35000514}.

\bibitem[{\citenamefont{Bj\"ork et~al.}(2001)\citenamefont{Bj\"ork, Jonsson,
  and S\'anchez~Soto}}]{Sanchez-Soto_2001}
\bibinfo{author}{\bibfnamefont{G.}~\bibnamefont{Bj\"ork}},
  \bibinfo{author}{\bibfnamefont{P.}~\bibnamefont{Jonsson}}, \bibnamefont{and}
  \bibinfo{author}{\bibfnamefont{L.~L.} \bibnamefont{S\'anchez~Soto}},
  \bibinfo{journal}{Phys. Rev. A} \textbf{\bibinfo{volume}{64}},
  \bibinfo{pages}{042106} (\bibinfo{year}{2001}),
  \urlprefix\url{https://link.aps.org/doi/10.1103/PhysRevA.64.042106}.

\bibitem[{\citenamefont{Y.~Hasegawa and Rauch}(2003)}]{NeutronBell2003}
\bibinfo{author}{\bibfnamefont{G.~B. M.~B.} \bibnamefont{Y.~Hasegawa},
  \bibfnamefont{R.~Loidl}} \bibnamefont{and}
  \bibinfo{author}{\bibfnamefont{H.}~\bibnamefont{Rauch}},
  \bibinfo{journal}{Nature} \textbf{\bibinfo{volume}{425}}, \bibinfo{pages}{45}
  (\bibinfo{year}{2003}), \urlprefix\url{https://doi.org/10.1038/nature01881}.

\bibitem[{\citenamefont{Matsukevich et~al.}(2008)\citenamefont{Matsukevich,
  Maunz, Moehring, Olmschenk, and Monroe}}]{Monroe2008}
\bibinfo{author}{\bibfnamefont{D.~N.} \bibnamefont{Matsukevich}},
  \bibinfo{author}{\bibfnamefont{P.}~\bibnamefont{Maunz}},
  \bibinfo{author}{\bibfnamefont{D.~L.} \bibnamefont{Moehring}},
  \bibinfo{author}{\bibfnamefont{S.}~\bibnamefont{Olmschenk}},
  \bibnamefont{and} \bibinfo{author}{\bibfnamefont{C.}~\bibnamefont{Monroe}},
  \bibinfo{journal}{Phys. Rev. Lett.} \textbf{\bibinfo{volume}{100}},
  \bibinfo{pages}{150404} (\bibinfo{year}{2008}),
  \urlprefix\url{https://link.aps.org/doi/10.1103/PhysRevLett.100.150404}.

\bibitem[{\citenamefont{Borges et~al.}(2010)\citenamefont{Borges, Hor-Meyll,
  Huguenin, and Khoury}}]{BellUFF2010}
\bibinfo{author}{\bibfnamefont{C.~V.~S.} \bibnamefont{Borges}},
  \bibinfo{author}{\bibfnamefont{M.}~\bibnamefont{Hor-Meyll}},
  \bibinfo{author}{\bibfnamefont{J.~A.~O.} \bibnamefont{Huguenin}},
  \bibnamefont{and} \bibinfo{author}{\bibfnamefont{A.~Z.}
  \bibnamefont{Khoury}}, \bibinfo{journal}{Phys. Rev. A}
  \textbf{\bibinfo{volume}{82}}, \bibinfo{pages}{033833}
  (\bibinfo{year}{2010}),
  \urlprefix\url{https://link.aps.org/doi/10.1103/PhysRevA.82.033833}.

\bibitem[{\citenamefont{Borges et~al.}(2012)\citenamefont{Borges, Khoury,
  Walborn, Ribeiro, Milman, and Keller}}]{BellOAM}
\bibinfo{author}{\bibfnamefont{C.~V.~S.} \bibnamefont{Borges}},
  \bibinfo{author}{\bibfnamefont{A.~Z.} \bibnamefont{Khoury}},
  \bibinfo{author}{\bibfnamefont{S.}~\bibnamefont{Walborn}},
  \bibinfo{author}{\bibfnamefont{P.~H.~S.} \bibnamefont{Ribeiro}},
  \bibinfo{author}{\bibfnamefont{P.}~\bibnamefont{Milman}}, \bibnamefont{and}
  \bibinfo{author}{\bibfnamefont{A.}~\bibnamefont{Keller}},
  \bibinfo{journal}{Phys. Rev. A} \textbf{\bibinfo{volume}{86}},
  \bibinfo{pages}{052107} (\bibinfo{year}{2012}),
  \urlprefix\url{https://link.aps.org/doi/10.1103/PhysRevA.86.052107}.

\bibitem[{\citenamefont{Pereira et~al.}(2014)\citenamefont{Pereira, Khoury, and
  Dechoum}}]{Leonardo2014}
\bibinfo{author}{\bibfnamefont{L.~J.} \bibnamefont{Pereira}},
  \bibinfo{author}{\bibfnamefont{A.~Z.} \bibnamefont{Khoury}},
  \bibnamefont{and} \bibinfo{author}{\bibfnamefont{K.}~\bibnamefont{Dechoum}},
  \bibinfo{journal}{Phys. Rev. A} \textbf{\bibinfo{volume}{90}},
  \bibinfo{pages}{053842} (\bibinfo{year}{2014}),
  \urlprefix\url{https://link.aps.org/doi/10.1103/PhysRevA.90.053842}.

\bibitem[{\citenamefont{Stoklasa et~al.}(2015)\citenamefont{Stoklasa, Motka,
  Rehacek, Hradil, S{\'{a}}nchez-Soto, and Agarwal}}]{Stoklasa_2015}
\bibinfo{author}{\bibfnamefont{B.}~\bibnamefont{Stoklasa}},
  \bibinfo{author}{\bibfnamefont{L.}~\bibnamefont{Motka}},
  \bibinfo{author}{\bibfnamefont{J.}~\bibnamefont{Rehacek}},
  \bibinfo{author}{\bibfnamefont{Z.}~\bibnamefont{Hradil}},
  \bibinfo{author}{\bibfnamefont{L.~L.} \bibnamefont{S{\'{a}}nchez-Soto}},
  \bibnamefont{and} \bibinfo{author}{\bibfnamefont{G.~S.}
  \bibnamefont{Agarwal}}, \bibinfo{journal}{New Journal of Physics}
  \textbf{\bibinfo{volume}{17}}, \bibinfo{pages}{113046}
  (\bibinfo{year}{2015}),
  \urlprefix\url{https://doi.org/10.1088%2F1367-2630%2F17%2F11%2F113046}.

\bibitem[{\citenamefont{Balthazar et~al.}(2016)\citenamefont{Balthazar, Souza,
  Caetano, Galv{\~a}o, Huguenin, and Khoury}}]{Balthazar2016}
\bibinfo{author}{\bibfnamefont{W.~F.} \bibnamefont{Balthazar}},
  \bibinfo{author}{\bibfnamefont{C.~E.~R.} \bibnamefont{Souza}},
  \bibinfo{author}{\bibfnamefont{D.~P.} \bibnamefont{Caetano}},
  \bibinfo{author}{\bibfnamefont{E.~F.} \bibnamefont{Galv{\~a}o}},
  \bibinfo{author}{\bibfnamefont{J.~A.~O.} \bibnamefont{Huguenin}},
  \bibnamefont{and} \bibinfo{author}{\bibfnamefont{A.~Z.}
  \bibnamefont{Khoury}}, \bibinfo{journal}{Opt. Lett.}
  \textbf{\bibinfo{volume}{41}}, \bibinfo{pages}{5797} (\bibinfo{year}{2016}),
  \urlprefix\url{http://ol.osa.org/abstract.cfm?URI=ol-41-24-5797}.

\bibitem[{\citenamefont{Rauch et~al.}(2018)\citenamefont{Rauch, Handsteiner,
  Hochrainer, Gallicchio, Friedman, Leung, Liu, Bulla, Ecker, Steinlechner
  et~al.}}]{Quasar2018}
\bibinfo{author}{\bibfnamefont{D.}~\bibnamefont{Rauch}},
  \bibinfo{author}{\bibfnamefont{J.}~\bibnamefont{Handsteiner}},
  \bibinfo{author}{\bibfnamefont{A.}~\bibnamefont{Hochrainer}},
  \bibinfo{author}{\bibfnamefont{J.}~\bibnamefont{Gallicchio}},
  \bibinfo{author}{\bibfnamefont{A.~S.} \bibnamefont{Friedman}},
  \bibinfo{author}{\bibfnamefont{C.}~\bibnamefont{Leung}},
  \bibinfo{author}{\bibfnamefont{B.}~\bibnamefont{Liu}},
  \bibinfo{author}{\bibfnamefont{L.}~\bibnamefont{Bulla}},
  \bibinfo{author}{\bibfnamefont{S.}~\bibnamefont{Ecker}},
  \bibinfo{author}{\bibfnamefont{F.}~\bibnamefont{Steinlechner}},
  \bibnamefont{et~al.}, \bibinfo{journal}{Phys. Rev. Lett.}
  \textbf{\bibinfo{volume}{121}}, \bibinfo{pages}{080403}
  (\bibinfo{year}{2018}),
  \urlprefix\url{https://link.aps.org/doi/10.1103/PhysRevLett.121.080403}.

\bibitem[{\citenamefont{Clauser et~al.}(1969)\citenamefont{Clauser, Horne,
  Shimony, and Holt}}]{CHSH}
\bibinfo{author}{\bibfnamefont{J.~F.} \bibnamefont{Clauser}},
  \bibinfo{author}{\bibfnamefont{M.~A.} \bibnamefont{Horne}},
  \bibinfo{author}{\bibfnamefont{A.}~\bibnamefont{Shimony}}, \bibnamefont{and}
  \bibinfo{author}{\bibfnamefont{R.~A.} \bibnamefont{Holt}},
  \bibinfo{journal}{Phys. Rev. Lett.} \textbf{\bibinfo{volume}{23}},
  \bibinfo{pages}{880} (\bibinfo{year}{1969}),
  \urlprefix\url{https://link.aps.org/doi/10.1103/PhysRevLett.23.880}.

\bibitem[{\citenamefont{Loudon}(1973)}]{Loudon1973}
\bibinfo{author}{\bibfnamefont{R.}~\bibnamefont{Loudon}},
  \emph{\bibinfo{title}{{The Quantum Theory of Light}}}
  (\bibinfo{publisher}{Clarendon Press}, \bibinfo{address}{Oxford},
  \bibinfo{year}{1973}), \urlprefix\url{https://cds.cern.ch/record/105699}.

\bibitem[{\citenamefont{Mandel and Wolf}(1995)}]{mandel_wolf_1995}
\bibinfo{author}{\bibfnamefont{L.}~\bibnamefont{Mandel}} \bibnamefont{and}
  \bibinfo{author}{\bibfnamefont{E.}~\bibnamefont{Wolf}},
  \emph{\bibinfo{title}{Optical Coherence and Quantum Optics}}
  (\bibinfo{publisher}{Cambridge University Press}, \bibinfo{year}{1995}).

\bibitem[{\citenamefont{{Dalibard, J.} et~al.}(1982)\citenamefont{{Dalibard,
  J.}, {Dupont-Roc, J.}, and {Cohen-Tannoudji, C.}}}]{Dalibard1982}
\bibinfo{author}{\bibnamefont{{Dalibard, J.}}},
  \bibinfo{author}{\bibnamefont{{Dupont-Roc, J.}}}, \bibnamefont{and}
  \bibinfo{author}{\bibnamefont{{Cohen-Tannoudji, C.}}}, \bibinfo{journal}{J.
  Phys. France} \textbf{\bibinfo{volume}{43}}, \bibinfo{pages}{1617}
  (\bibinfo{year}{1982}),
  \urlprefix\url{https://doi.org/10.1051/jphys:0198200430110161700}.

\bibitem[{\citenamefont{De~Bi\`evre et~al.}(2019)\citenamefont{De~Bi\`evre,
  Horoshko, Patera, and Kolobov}}]{Kolobov2019}
\bibinfo{author}{\bibfnamefont{S.}~\bibnamefont{De~Bi\`evre}},
  \bibinfo{author}{\bibfnamefont{D.~B.} \bibnamefont{Horoshko}},
  \bibinfo{author}{\bibfnamefont{G.}~\bibnamefont{Patera}}, \bibnamefont{and}
  \bibinfo{author}{\bibfnamefont{M.~I.} \bibnamefont{Kolobov}},
  \bibinfo{journal}{Phys. Rev. Lett.} \textbf{\bibinfo{volume}{122}},
  \bibinfo{pages}{080402} (\bibinfo{year}{2019}),
  \urlprefix\url{https://link.aps.org/doi/10.1103/PhysRevLett.122.080402}.

\bibitem[{\citenamefont{Casado et~al.}(1998)\citenamefont{Casado, Marshall, and
  Santos}}]{Casado1998}
\bibinfo{author}{\bibfnamefont{A.}~\bibnamefont{Casado}},
  \bibinfo{author}{\bibfnamefont{T.~W.} \bibnamefont{Marshall}},
  \bibnamefont{and} \bibinfo{author}{\bibfnamefont{E.}~\bibnamefont{Santos}},
  \bibinfo{journal}{J. Opt. Soc. Am. B} \textbf{\bibinfo{volume}{15}},
  \bibinfo{pages}{1572} (\bibinfo{year}{1998}),
  \urlprefix\url{http://josab.osa.org/abstract.cfm?URI=josab-15-5-1572}.

\bibitem[{\citenamefont{Olsen et~al.}(2001)\citenamefont{Olsen, Dechoum, and
  Plimak}}]{Olsen2001}
\bibinfo{author}{\bibfnamefont{M.}~\bibnamefont{Olsen}},
  \bibinfo{author}{\bibfnamefont{K.}~\bibnamefont{Dechoum}}, \bibnamefont{and}
  \bibinfo{author}{\bibfnamefont{L.}~\bibnamefont{Plimak}},
  \bibinfo{journal}{Optics Communications} \textbf{\bibinfo{volume}{190}},
  \bibinfo{pages}{261 } (\bibinfo{year}{2001}), ISSN \bibinfo{issn}{0030-4018},
  \urlprefix\url{http://www.sciencedirect.com/science/article/pii/S0030401801011002}.

\bibitem[{\citenamefont{da~Silva et~al.}(2008)\citenamefont{da~Silva, Khoury,
  and Dechoum}}]{Luciano2008}
\bibinfo{author}{\bibfnamefont{L.~F.} \bibnamefont{da~Silva}},
  \bibinfo{author}{\bibfnamefont{A.~Z.} \bibnamefont{Khoury}},
  \bibnamefont{and} \bibinfo{author}{\bibfnamefont{K.}~\bibnamefont{Dechoum}},
  \bibinfo{journal}{Phys. Rev. A} \textbf{\bibinfo{volume}{78}},
  \bibinfo{pages}{025803} (\bibinfo{year}{2008}),
  \urlprefix\url{https://link.aps.org/doi/10.1103/PhysRevA.78.025803}.

\bibitem[{\citenamefont{Casado et~al.}(2019)\citenamefont{Casado, Guerra, and
  Plácido}}]{Casado_2019}
\bibinfo{author}{\bibfnamefont{A.}~\bibnamefont{Casado}},
  \bibinfo{author}{\bibfnamefont{S.}~\bibnamefont{Guerra}}, \bibnamefont{and}
  \bibinfo{author}{\bibfnamefont{J.}~\bibnamefont{Plácido}},
  \bibinfo{journal}{Atoms} \textbf{\bibinfo{volume}{7}}, \bibinfo{pages}{76}
  (\bibinfo{year}{2019}), ISSN \bibinfo{issn}{2218-2004},
  \urlprefix\url{http://dx.doi.org/10.3390/atoms7030076}.

\bibitem[{\citenamefont{Eberly}(2015)}]{Eberly2015}
\bibinfo{author}{\bibfnamefont{J.}~\bibnamefont{Eberly}},
  \bibinfo{journal}{Contemporary Physics} \textbf{\bibinfo{volume}{56}},
  \bibinfo{pages}{407} (\bibinfo{year}{2015}),
  \eprint{https://doi.org/10.1080/00107514.2015.1080949},
  \urlprefix\url{https://doi.org/10.1080/00107514.2015.1080949}.

\bibitem[{\citenamefont{Eberly et~al.}(2016)\citenamefont{Eberly, Qian, Asma
  Al~Qasimi, Ali, Alonso, Guti{\'{e}}rrez-Cuevas, Little, Howell, Malhotra, and
  Vamivakas}}]{Eberly2016}
\bibinfo{author}{\bibfnamefont{J.~H.} \bibnamefont{Eberly}},
  \bibinfo{author}{\bibfnamefont{X.~F.} \bibnamefont{Qian}},
  \bibinfo{author}{\bibfnamefont{A.}~\bibnamefont{Asma Al~Qasimi}},
  \bibinfo{author}{\bibfnamefont{H.}~\bibnamefont{Ali}},
  \bibinfo{author}{\bibfnamefont{M.~A.} \bibnamefont{Alonso}},
  \bibinfo{author}{\bibfnamefont{R.}~\bibnamefont{Guti{\'{e}}rrez-Cuevas}},
  \bibinfo{author}{\bibfnamefont{B.~J.} \bibnamefont{Little}},
  \bibinfo{author}{\bibfnamefont{J.~C.} \bibnamefont{Howell}},
  \bibinfo{author}{\bibfnamefont{T.}~\bibnamefont{Malhotra}}, \bibnamefont{and}
  \bibinfo{author}{\bibfnamefont{A.~N.} \bibnamefont{Vamivakas}},
  \bibinfo{journal}{Physica Scripta} \textbf{\bibinfo{volume}{91}},
  \bibinfo{pages}{063003} (\bibinfo{year}{2016}),
  \urlprefix\url{https://doi.org/10.1088%2F0031-8949%2F91%2F6%2F063003}.

\bibitem[{\citenamefont{Qian et~al.}(2016)\citenamefont{Qian, Malhotra,
  Vamivakas, and Eberly}}]{Qian2016}
\bibinfo{author}{\bibfnamefont{X.~F.} \bibnamefont{Qian}},
  \bibinfo{author}{\bibfnamefont{T.}~\bibnamefont{Malhotra}},
  \bibinfo{author}{\bibfnamefont{A.~N.} \bibnamefont{Vamivakas}},
  \bibnamefont{and} \bibinfo{author}{\bibfnamefont{J.~H.}
  \bibnamefont{Eberly}}, \bibinfo{journal}{Phys. Rev. Lett.}
  \textbf{\bibinfo{volume}{117}}, \bibinfo{pages}{153901}
  (\bibinfo{year}{2016}),
  \urlprefix\url{https://link.aps.org/doi/10.1103/PhysRevLett.117.153901}.

\bibitem[{\citenamefont{Eberly et~al.}(2017)\citenamefont{Eberly, Qian, and
  Vamivakas}}]{Eberly2017}
\bibinfo{author}{\bibfnamefont{J.~H.} \bibnamefont{Eberly}},
  \bibinfo{author}{\bibfnamefont{X.-F.} \bibnamefont{Qian}}, \bibnamefont{and}
  \bibinfo{author}{\bibfnamefont{A.~N.} \bibnamefont{Vamivakas}},
  \bibinfo{journal}{Optica} \textbf{\bibinfo{volume}{4}}, \bibinfo{pages}{1113}
  (\bibinfo{year}{2017}),
  \urlprefix\url{http://www.osapublishing.org/optica/abstract.cfm?URI=optica-4-9-1113}.

\bibitem[{\citenamefont{Zela}(2018)}]{DeZela2018}
\bibinfo{author}{\bibfnamefont{F.~D.} \bibnamefont{Zela}},
  \bibinfo{journal}{Optica} \textbf{\bibinfo{volume}{5}}, \bibinfo{pages}{243}
  (\bibinfo{year}{2018}),
  \urlprefix\url{http://www.osapublishing.org/optica/abstract.cfm?URI=optica-5-3-243}.

\bibitem[{\citenamefont{Gonzales et~al.}(2018)\citenamefont{Gonzales,
  S{\'{a}}nchez, Barberena, Yugra, Caballero, and Zela}}]{Gonzales2018}
\bibinfo{author}{\bibfnamefont{J.}~\bibnamefont{Gonzales}},
  \bibinfo{author}{\bibfnamefont{P.}~\bibnamefont{S{\'{a}}nchez}},
  \bibinfo{author}{\bibfnamefont{D.}~\bibnamefont{Barberena}},
  \bibinfo{author}{\bibfnamefont{Y.}~\bibnamefont{Yugra}},
  \bibinfo{author}{\bibfnamefont{R.}~\bibnamefont{Caballero}},
  \bibnamefont{and} \bibinfo{author}{\bibfnamefont{F.~D.} \bibnamefont{Zela}},
  \bibinfo{journal}{Journal of Physics B: Atomic, Molecular and Optical
  Physics} \textbf{\bibinfo{volume}{51}}, \bibinfo{pages}{045401}
  (\bibinfo{year}{2018}),
  \urlprefix\url{https://doi.org/10.1088%2F1361-6455%2Faaa185}.

\bibitem[{\citenamefont{Kwiat et~al.}(1995)\citenamefont{Kwiat, Mattle,
  Weinfurter, Zeilinger, Sergienko, and Shih}}]{Kwiat1995}
\bibinfo{author}{\bibfnamefont{P.~G.} \bibnamefont{Kwiat}},
  \bibinfo{author}{\bibfnamefont{K.}~\bibnamefont{Mattle}},
  \bibinfo{author}{\bibfnamefont{H.}~\bibnamefont{Weinfurter}},
  \bibinfo{author}{\bibfnamefont{A.}~\bibnamefont{Zeilinger}},
  \bibinfo{author}{\bibfnamefont{A.~V.} \bibnamefont{Sergienko}},
  \bibnamefont{and} \bibinfo{author}{\bibfnamefont{Y.}~\bibnamefont{Shih}},
  \bibinfo{journal}{Phys. Rev. Lett.} \textbf{\bibinfo{volume}{75}},
  \bibinfo{pages}{4337} (\bibinfo{year}{1995}),
  \urlprefix\url{https://link.aps.org/doi/10.1103/PhysRevLett.75.4337}.

\bibitem[{\citenamefont{Santos et~al.}(2001)\citenamefont{Santos, Milman,
  Khoury, and Ribeiro}}]{Santos2001}
\bibinfo{author}{\bibfnamefont{M.~F.} \bibnamefont{Santos}},
  \bibinfo{author}{\bibfnamefont{P.}~\bibnamefont{Milman}},
  \bibinfo{author}{\bibfnamefont{A.~Z.} \bibnamefont{Khoury}},
  \bibnamefont{and} \bibinfo{author}{\bibfnamefont{P.~H.~S.}
  \bibnamefont{Ribeiro}}, \bibinfo{journal}{Physical Review A}
  \textbf{\bibinfo{volume}{64}}, \bibinfo{pages}{023804}
  (\bibinfo{year}{2001}).

\bibitem[{\citenamefont{Caetano et~al.}(2003)\citenamefont{Caetano, Ribeiro,
  Pardal, and Khoury}}]{Juliana2003}
\bibinfo{author}{\bibfnamefont{D.~P.} \bibnamefont{Caetano}},
  \bibinfo{author}{\bibfnamefont{P.~H.~S.} \bibnamefont{Ribeiro}},
  \bibinfo{author}{\bibfnamefont{J.~T.~C.} \bibnamefont{Pardal}},
  \bibnamefont{and} \bibinfo{author}{\bibfnamefont{A.~Z.}
  \bibnamefont{Khoury}}, \bibinfo{journal}{Phys. Rev. A}
  \textbf{\bibinfo{volume}{68}}, \bibinfo{pages}{023805}
  (\bibinfo{year}{2003}),
  \urlprefix\url{https://link.aps.org/doi/10.1103/PhysRevA.68.023805}.

\bibitem[{\citenamefont{Carmichael}(1999)}]{carmichael_1999}
\bibinfo{author}{\bibfnamefont{H.}~\bibnamefont{Carmichael}},
  \emph{\bibinfo{title}{Statistical Methods in Quantum Optics 1}}
  (\bibinfo{publisher}{Springer}, \bibinfo{year}{1999}).

\bibitem[{\citenamefont{Dononov}(2003)}]{Dodonov_2003}
\bibinfo{author}{\bibfnamefont{M.~V.~I.} \bibnamefont{Dononov},
  \bibfnamefont{V.~V.}}, \emph{\bibinfo{title}{Theory of Nonclassical States of
  Light}} (\bibinfo{publisher}{Taylor and Francis}, \bibinfo{year}{2003}).

\bibitem[{\citenamefont{Dechoum et~al.}(2010)\citenamefont{Dechoum, Hahn,
  Vallejos, and Khoury}}]{WignerUFF2010}
\bibinfo{author}{\bibfnamefont{K.}~\bibnamefont{Dechoum}},
  \bibinfo{author}{\bibfnamefont{M.~D.} \bibnamefont{Hahn}},
  \bibinfo{author}{\bibfnamefont{R.~O.} \bibnamefont{Vallejos}},
  \bibnamefont{and} \bibinfo{author}{\bibfnamefont{A.~Z.}
  \bibnamefont{Khoury}}, \bibinfo{journal}{Phys. Rev. A}
  \textbf{\bibinfo{volume}{81}}, \bibinfo{pages}{043834}
  (\bibinfo{year}{2010}),
  \urlprefix\url{https://link.aps.org/doi/10.1103/PhysRevA.81.043834}.

\bibitem[{\citenamefont{Stiller et~al.}(2017)\citenamefont{Stiller, Seyfarth,
  and Leuchs}}]{LesHouches2013}
\bibinfo{author}{\bibfnamefont{B.}~\bibnamefont{Stiller}},
  \bibinfo{author}{\bibfnamefont{U.}~\bibnamefont{Seyfarth}}, \bibnamefont{and}
  \bibinfo{author}{\bibfnamefont{G.}~\bibnamefont{Leuchs}},
  \emph{\bibinfo{title}{{Les Houches 2013: Quantum Optics and Nanophotonics
  (Chapter 4)}}} (\bibinfo{publisher}{Oxford University Press},
  \bibinfo{address}{Oxford, UK}, \bibinfo{year}{2017}),
  \urlprefix\url{https://global.oup.com/academic}.

\bibitem[{\citenamefont{Stiller et~al.}(2014)\citenamefont{Stiller, Seyfarth,
  and Leuchs}}]{Stiller2014}
\bibinfo{author}{\bibfnamefont{B.}~\bibnamefont{Stiller}},
  \bibinfo{author}{\bibfnamefont{U.}~\bibnamefont{Seyfarth}}, \bibnamefont{and}
  \bibinfo{author}{\bibfnamefont{G.}~\bibnamefont{Leuchs}},
  \emph{\bibinfo{title}{Temporal and spectral properties of quantum light}}
  (\bibinfo{year}{2014}), \eprint{1411.3765}.

\bibitem[{\citenamefont{Thearle et~al.}(2018)\citenamefont{Thearle, Janousek,
  Armstrong, Hosseini, Sch\"unemann~(Mraz), Assad, Symul, James, Huntington,
  Ralph et~al.}}]{PingKoyLam2018}
\bibinfo{author}{\bibfnamefont{O.}~\bibnamefont{Thearle}},
  \bibinfo{author}{\bibfnamefont{J.}~\bibnamefont{Janousek}},
  \bibinfo{author}{\bibfnamefont{S.}~\bibnamefont{Armstrong}},
  \bibinfo{author}{\bibfnamefont{S.}~\bibnamefont{Hosseini}},
  \bibinfo{author}{\bibfnamefont{M.}~\bibnamefont{Sch\"unemann~(Mraz)}},
  \bibinfo{author}{\bibfnamefont{S.}~\bibnamefont{Assad}},
  \bibinfo{author}{\bibfnamefont{T.}~\bibnamefont{Symul}},
  \bibinfo{author}{\bibfnamefont{M.~R.} \bibnamefont{James}},
  \bibinfo{author}{\bibfnamefont{E.}~\bibnamefont{Huntington}},
  \bibinfo{author}{\bibfnamefont{T.~C.} \bibnamefont{Ralph}},
  \bibnamefont{et~al.}, \bibinfo{journal}{Phys. Rev. Lett.}
  \textbf{\bibinfo{volume}{120}}, \bibinfo{pages}{040406}
  (\bibinfo{year}{2018}),
  \urlprefix\url{https://link.aps.org/doi/10.1103/PhysRevLett.120.040406}.

\bibitem[{\citenamefont{Milonni}(1994)}]{Milonni1994}
\bibinfo{author}{\bibfnamefont{P.}~\bibnamefont{Milonni}},
  \emph{\bibinfo{title}{{The Quantum Vacuum. An Introduction to Quantum
  Electrodynamics}}} (\bibinfo{publisher}{Academic Press},
  \bibinfo{address}{Cambridge, Massachusetts}, \bibinfo{year}{1994}),
  \urlprefix\url{https://www.elsevier.com/books/the-quantum-vacuum/milonni/978-0-08-057149-2}.

\bibitem[{\citenamefont{Benea-Chelmus et~al.}(2019)\citenamefont{Benea-Chelmus,
  Settembrini, Scalari, and Faist}}]{Vacuum-measure}
\bibinfo{author}{\bibfnamefont{I.~C.} \bibnamefont{Benea-Chelmus}},
  \bibinfo{author}{\bibfnamefont{F.~F.} \bibnamefont{Settembrini}},
  \bibinfo{author}{\bibfnamefont{G.}~\bibnamefont{Scalari}}, \bibnamefont{and}
  \bibinfo{author}{\bibfnamefont{J.}~\bibnamefont{Faist}},
  \bibinfo{journal}{Nature} \textbf{\bibinfo{volume}{568}},
  \bibinfo{pages}{202} (\bibinfo{year}{2019}),
  \urlprefix\url{https://doi.org/10.1038/s41586-019-1083-9}.

\bibitem[{\citenamefont{Cassemiro et~al.}(2007)\citenamefont{Cassemiro, Villar,
  Martinelli, and Nussenzveig}}]{Cassemiro:07}
\bibinfo{author}{\bibfnamefont{K.~N.} \bibnamefont{Cassemiro}},
  \bibinfo{author}{\bibfnamefont{A.~S.} \bibnamefont{Villar}},
  \bibinfo{author}{\bibfnamefont{M.}~\bibnamefont{Martinelli}},
  \bibnamefont{and}
  \bibinfo{author}{\bibfnamefont{P.}~\bibnamefont{Nussenzveig}},
  \bibinfo{journal}{Opt. Express} \textbf{\bibinfo{volume}{15}},
  \bibinfo{pages}{18236} (\bibinfo{year}{2007}),
  \urlprefix\url{http://www.opticsexpress.org/abstract.cfm?URI=oe-15-26-18236}.

\bibitem[{\citenamefont{Coelho et~al.}(2009)\citenamefont{Coelho, Barbosa,
  Cassemiro, Villar, Martinelli, and Nussenzveig}}]{Coelho2009}
\bibinfo{author}{\bibfnamefont{A.~S.} \bibnamefont{Coelho}},
  \bibinfo{author}{\bibfnamefont{F.~A.~S.} \bibnamefont{Barbosa}},
  \bibinfo{author}{\bibfnamefont{K.~N.} \bibnamefont{Cassemiro}},
  \bibinfo{author}{\bibfnamefont{A.~S.} \bibnamefont{Villar}},
  \bibinfo{author}{\bibfnamefont{M.}~\bibnamefont{Martinelli}},
  \bibnamefont{and}
  \bibinfo{author}{\bibfnamefont{P.}~\bibnamefont{Nussenzveig}},
  \bibinfo{journal}{Science} \textbf{\bibinfo{volume}{326}},
  \bibinfo{pages}{823} (\bibinfo{year}{2009}), ISSN \bibinfo{issn}{0036-8075},
  \urlprefix\url{https://science.sciencemag.org/content/326/5954/823}.

\bibitem[{\citenamefont{McCormick et~al.}(2008)\citenamefont{McCormick, Marino,
  Boyer, and Lett}}]{Lett2008}
\bibinfo{author}{\bibfnamefont{C.~F.} \bibnamefont{McCormick}},
  \bibinfo{author}{\bibfnamefont{A.~M.} \bibnamefont{Marino}},
  \bibinfo{author}{\bibfnamefont{V.}~\bibnamefont{Boyer}}, \bibnamefont{and}
  \bibinfo{author}{\bibfnamefont{P.~D.} \bibnamefont{Lett}},
  \bibinfo{journal}{Phys. Rev. A} \textbf{\bibinfo{volume}{78}},
  \bibinfo{pages}{043816} (\bibinfo{year}{2008}),
  \urlprefix\url{https://link.aps.org/doi/10.1103/PhysRevA.78.043816}.

\bibitem[{\citenamefont{Boyer et~al.}(2008)\citenamefont{Boyer, Marino, Pooser,
  and Lett}}]{Boyer2008}
\bibinfo{author}{\bibfnamefont{V.}~\bibnamefont{Boyer}},
  \bibinfo{author}{\bibfnamefont{A.~M.} \bibnamefont{Marino}},
  \bibinfo{author}{\bibfnamefont{R.~C.} \bibnamefont{Pooser}},
  \bibnamefont{and} \bibinfo{author}{\bibfnamefont{P.~D.} \bibnamefont{Lett}},
  \bibinfo{journal}{Science} \textbf{\bibinfo{volume}{321}},
  \bibinfo{pages}{544} (\bibinfo{year}{2008}), ISSN \bibinfo{issn}{0036-8075},
  \urlprefix\url{https://science.sciencemag.org/content/321/5888/544}.

\bibitem[{\citenamefont{Ferraz et~al.}(2005)\citenamefont{Ferraz, Felinto,
  Acioli, and Vianna}}]{Ferraz2005}
\bibinfo{author}{\bibfnamefont{J.}~\bibnamefont{Ferraz}},
  \bibinfo{author}{\bibfnamefont{D.}~\bibnamefont{Felinto}},
  \bibinfo{author}{\bibfnamefont{L.~H.} \bibnamefont{Acioli}},
  \bibnamefont{and} \bibinfo{author}{\bibfnamefont{S.~S.}
  \bibnamefont{Vianna}}, \bibinfo{journal}{Opt. Lett.}
  \textbf{\bibinfo{volume}{30}}, \bibinfo{pages}{1876} (\bibinfo{year}{2005}),
  \urlprefix\url{http://ol.osa.org/abstract.cfm?URI=ol-30-14-1876}.

\end{thebibliography}

\end{document}